\documentclass{article}
\begin{document}
\newcommand{\be}{\begin{equation}}
\newcommand{\p}{\partial}
\newcommand{\nn}{\nonumber}

\newcommand{\ee}{\end{equation}}
\newcommand{\beqn}{\begin{eqnarray}}
\newcommand{\eeqn}{\end{eqnarray}}
\newcommand{\f}{\frac}
\newcommand{\w}{\wedge}

\begin{titlepage}
\flushright{WIS/2/05-JAN-DPP}
\flushright{hep-th/0501012}

\vspace{1in}

\begin{center}
\Large
{\bf A new Ricci flat geometry }
\vspace{1in}

\normalsize

\large{ Shesansu Sekhar  Pal }\\

\normalsize
\vspace{.7in}

{\em Weizmann institute of  science,\\
76100 Rehovot, Israel }\\
{\sf shesansu.pal@weizmann.ac.il}
\end{center}

\vspace{1in}

\baselineskip=24pt
\begin{abstract}
We are proposing a new Ricci flat metric constructed from 
an infinite family of Sasaki-Einstein, $Y^{(p,q)}$, geometries. This 
geometry contains a free parameter $s$ and in the $s\rightarrow 0$ limit
 we get back the usual CY.
When this geometry is probed both by a stack of D3 and 
fractional D3 branes  then the   
 corresponding supergravity solution is found which is a warped product 
of this new
6-dimensional geometry and the flat $R^{3,1}$. This solution in the specific 
limit as mentioned above reproduces the solution found in hep-th/0412193.
The integrated five-form field strength 
over  $S^2\times S^3$ goes logarithmically but the  argument of Log function
is different than has been found before.


\end{abstract}

\end{titlepage}

\section{Introduction}

The study of AdS/CFT \cite{jm}\footnote{ For a review \cite{agmoo}.} has 
been instrumental in  sharpening our understanding of 
various gravitational theories with singularity, especially branes probing
various conical singularities. The famous example before \cite{gmsw} is the
conifold where probing of the tip of conifold with various D3-branes provides   us the $AdS_5\times T^{1,1}$ space which in the corresponding dual 
field theory \cite{kw}  would become 
conformal or nonconformal depending on the absence or presence of fractional 
D3-branes. The latter are D5 branes wrapped on the 2-cycles of the conifold.
More interestingly, these field theories preserve four supercharges in four
spacetime dimensions i.e. it preserves ${\cal N}=1$ supersymmetry.   

There has been an obstacle to compute the scaling dimensions or the
anomalous dimensions of the chiral 
superfields that appear in these supersymmetric field theories due to the
appearance of various R-symmetries. In \cite{iw}, it has been suggested 
to consider the R-currents which maximizes the conformal anomaly, a, over all 
the possible R-currents which appear in the superconformal multiplet.  There 
has been interesting developments on the a-maximization principle in 
\cite{dk, biww}. 

Recently, in \cite{gmsw} an infinite family of Sasaki-Einstein manifold 
 $Y^{(p,q)}$ has 
been constructed by studying supersymmetric AdS solutions in 11-dimensional 
supergravity \cite{gmsw1}. These geometries are described by two positive
integers p and q with a
restriction $q < p$. The topology of this space is $S^2\times S^3$ and have 
the isometry group $SU(2)\times U(1)^2$. It has been shown in 
\cite{gmsw,ms} that there exists 
a Killing vector called Reeb vector which in the dual field theory is 
isomorphic to the R-symmetry, which is $\sim  \f{\p}{\p\psi}$ and depending 
on the orbits of this Killing vector field one gets regular, quasi-regular 
and  irregular SE manifolds. This newly found infinite family of 
Sasaki-Einstein geometries falls into
the last class i.e. of irregular type. One of the intrinsic property of 
the dual field theory associated to this kind irregular geometry is that
their central charges are irrational and hence the volumes of these geometries
are irrational too.
As we 
have mentioned already in the previous paragraph the dual field theory is
an ${\cal N}=1$ superconformal field theory when there is no fractional 
D3 branes otherwise it will be non-conformal field theory like \cite{ks}
with the same amount of supersymmetry. These conformal field theories 
has been studied in \cite{ms, bbc, bfhms} and has been shown to possess
Seiberg duality \cite{ns}.  

In a related development in \cite{ehk} found the supergravity solution
by probing these singular Calabi-Yau's with a stack of both D3 and 
fractional D3 branes.
These solution indeed shows the characteristic feature of Seiberg duality i.e.
the integrated five form field strength over $S^2\times S^3$ goes as $log~r$.

In this paper we are proposing a new 6-dimensional Ricci-flat geometry unlike
the usual--  a cone over  $Y^{(p,q)}$ i.e. eq.(\ref{cy}). This new Ricci flat
geometry contains a free parameter $s$ and in the limit of $s \rightarrow 0$
it gives back the eq. (\ref{cy}). We also present the supergravity 
solution for these Ricci flat geometry and in the above mentioned limit we
get back the solution presented in \cite{ehk}. An interesting point to note
that the warp factor depends on both the radial coordinate and the angular
coordinate $y$, which has been mentioned in \cite{ehk}. But, for our choice
of metric ansatz eq.(\ref{metric_10}), it follows that the y dependent part
of $h$ is universal i.e its form is same irrespective of the form of $G_1$
and $G_2$, the $H(y)$ of eq.(\ref{H(y)}).

The appearance of the parameter $s$ in the geometry corresponds to a closed
string moduli and it would be interesting to understand whether this moduli 
corresponds to a normalisable or non normalisable mode, \footnote{I would like
to thank M. Berkooz for suggesting to study this.} and the presence or absence
 of singularity at $\rho=0$, which we will not do 
that here but will do that in our future studies.

\section{The new geometry}

The new six dimensional Ricci flat  geometry that we are proposing is guessed 
from the existing 5-dimensional $Y^{(p,q)}$ Sasaki-Einstein metric. Its form 
look like

\be 
\label{geometry}
ds^2_6=K^{-1}(\rho) d\rho^2+K(\rho) \rho^2 {e^{\psi}}^2+ (\rho^2+s^2) ({e^{\theta}}^2+{e^{\phi}}^2+{e^y}^2+{e^{\beta}}^2),
\ee
where the one forms are defined as
\beqn
& &e^{\theta}=\sqrt{\f{1-c y}{6}}d\theta,\quad e^{\phi}=\sqrt{\f{1-c y}{6}}sin\theta
d\phi,\nn \\ & &
e^y=\f{1}{\sqrt{w(y)v(y)}} dy, \quad e^{\beta}= \f{\sqrt{w(y)v(y)}}{6}(d\beta+c~ cos\theta d\phi), \nn \\& &
e^{\psi}=\f{1}{3}[d\psi-cos\theta d\phi+y(d\beta+c ~cos\theta d\phi)]
\eeqn 
and \footnote{ We have written v(y) as opposed to  q(y), following 
\cite{ehk}, may be to avoid the confusion of q appearing in $Y^{(p,q)}$ and 
in the geometry.}
\beqn
& &K(\rho)=\f{\rho^4+3 s^2 \rho^2+3 s^4}{(\rho^2+s^2)^2},\nn \\& &
  w(y)= \f{2(b-y^2)}{(1-c y)}, \quad v(y)=\f{b-3y^2+2cy^3}{b-y^2}
\eeqn
As discussed in \cite{gmsw} that one can rescale the coordinate y and the 
parameter $c$ can be set to unity i.e. $c=1$. From now onwards we shall follow this except towards the end of this section. 
The above solution  makes sense only when the y coordinate stays 
between the two smallest roots of $b-3y^2+2y^3$ and are given by
\beqn
y_1&=&\f{1}{4p}\bigg(2p-3q-\sqrt{4p^2-3q^2}\bigg),\nn \\
y_2&=&\f{1}{4p}\bigg(2p+3q-\sqrt{4p^2-3q^2}\bigg),
\eeqn
where 
\be
b=\f{1}{2}-\f{p^2-3q^2}{4p^3}\sqrt{4p^2-3q^2}.
\ee
Following \cite{gmsw}, one can define $\alpha=-\beta/6-\psi/6$ and  
re express 
the metric written above in a different from and for our purpose the exact form that is not important. The aim of introducing the coordinate $\alpha$ is to 
\footnote{ In \cite{gmsw}, they first wrote down their metric in $(\theta, 
\phi, y, \beta, \alpha)$ coordinates.} mention that this coordinate has a 
period
of $2\pi\ell$. So, the ranges of various coordinates are: 
$0\leq\theta\leq \pi, 0\leq\phi\leq 2\pi, 0\leq\psi\leq 2\pi, 
y_1\leq y\leq y_2$ and $0\leq \alpha\leq 2\pi\ell$.

The space $Y^{(p,q)}$ has two independent parameters, corresponding to the 
two Chern numbers and the period of $\alpha$ and the volume of this space
depends on these two parameters $(p,q)$.  
\beqn
& &\ell=\f{q}{3 q^2-2p^2+p\sqrt{4 p^2-3 q^2}}\nn \\
& &{\rm Vol[Y^{(p,q)}]}=\f{q^2[2p+\sqrt{4p^2-3q^2}]}{3p^2[3q^2-2p^2+
p\sqrt{4p^2-3q^2}]},
\eeqn
with a restriction of $p >q$. 

It is easy to note that in the $s\rightarrow 0$ limit we do get back the 
usual way of constructing Calabi-Yau  from the 5-dimensional $Y^{(p,q)}$ 
geometry i.e.
\be 
\label{cy}
ds^2=dr^2+r^2ds^2_{Y^{(p,q)}}. 
\ee

Let us take  the form of $K(\rho)$ that appear in eq.(\ref{geometry}) as
\be
\label{gen_k}
K(\rho)=\f{c_1+\rho^6+3\rho^4s^2+3s^4\rho^2}{\rho^2(\rho^2+s^2)^2},
\ee 
where $c_1$ is a constant\footnote{We shall work
in the $c_1=0$ limit in next section.}. This metric is also  Ricci-flat in fact it is 
related to the earlier $K(\rho)$ by some change of coordinates and defining 
new constants. The 
interesting thing to note that in the limit of setting $c_1=s^6$, we
do get back our singular geometry i.e. eq.(\ref{cy}). The most interesting 
point  is that this geometry is a Calabi-Yau. Its Kahlerian behavior 
is shown in \cite{pc}. 

Let us recall  the geometry of the resolved conifold metric from \cite{pt}
\be
\label{resolved_coni}
ds^2={\tilde K}^{-1}(r)dr^2+{\tilde K}(r)r^2{e^{\psi}}^2+(r^2+s^2_1)({e^{\theta_1}}^2+{e^{\phi_1}}^2)+(r^2+s^2_2)({e^{\theta_2}}^2+{e^{\phi_2}}^2),
\ee
where 
${\tilde K}(r)=\f{c_2+r^6+3/2(s^2_1+ s^2_2)r^2+3r^2s^2_1s^2_2}
{r^2(r^2+s^2_1)(r^2+s^2_2)}$ with $c_2=\f{s^4_1}{2}(3s^2_2-s^2_1)$. 

It is easy to see that in the $c=0$ limit of eq.(\ref{geometry})
with eq.(\ref{gen_k}) gives the same metric as eq.(\ref{resolved_coni})
 with $s_1=s_2$ and ignoring the exact form of $c_2$. From this it is
tempting to think that eq.(\ref{geometry}) with   eq.(\ref{gen_k}) may
give hints to the geometry of resolved \footnote {More on it latter.}
$Y^{p,q}$ geometry.  

\section{The solution}
We shall find the  solution for the geometry written in 
eq.(\ref{geometry}) by first constructing a closed 3-form field strength
which obeys
the imaginary-self-duality (ISD) condition with respect to the geometry 
given in eq.(\ref{geometry}). We believe that this is enough to show that
the supergravity solution  preserve ${\cal N}=1$ supersymmetry. In \cite{ak},
it has been shown whenever the 3-form field strength becomes (2,1) and 
obeys the ISD condition and more importantly, for the case when the warp 
factor $h(\tau)$ is a function only of the radial coordinate $\tau$  then it 
necessarily preserves supersymmetry. However, for  the irregular SE geometries
the warp factor is a function of both the radial and one of the angular 
coordinate, $y$. So, it is important to check that this solution preserves 
 supersymmetry \footnote{which we will not do it here.}.   
For the form of our 6-dimensional metric  
we shall get the solution with the following ansatz to metric.
 

The ansatz to the 10-dimensional geometry is
\be
\label{metric_10}
ds^2=h^{-1/2}ds^2_4+h^{1/2}\bigg( G^2_1(\tau)[d\tau^2+{e^{\psi}}^2]+G^2_2(\tau)
[{e^{\theta}}^2+{e^{\phi}}^2+{e^y}^2+{ e^{\beta}}^2]\bigg).
\ee
For $G_1=G_2=r$ and $\tau=\ln r$, we get back the 6-dimensional 
geometry  eq.(\ref{cy}) and the solution derived in \cite{ehk}. Whereas in 
our case 
\be
\label{solution}
G_1=\sqrt{K(\rho)}\rho,~ G_2=\sqrt{\rho^2+s^2},~ \tau=\f{1}{6}\ln [\rho^6+3 s^2\rho^4+3 s^4\rho^2]. 
\ee

The dilaton, $\phi$ is \footnote{The dilaton should not be confused with the 
angular coordinate appear in the geometry.},  assumed to be constant and 
the axion, $C$, is set to zero i.e.
\be
e^{\phi}\equiv e^{\phi_0}=g_s, \quad C=0.
\ee

The NS-NS 2-form potential, $B_2$,  has the form
\be
B_2=g_sMKK^{\prime} f(\tau) F(y)[e^{\theta}\wedge e^{\phi}-e^y\wedge e^{\beta}]
\ee 
with the form of  F(y) as
\be
F(y)=\f{1}{(1-y)^2}.
\ee
The corresponding 3-form field strength, $H_3$ and RR 3-form field strength, 
$F_3$ is
\beqn
H_3&=&g_sMKK^{\prime} F(y)\f{df(\tau)}{d\tau}d\tau\wedge \bigg(e^{\theta}\wedge e^{\phi}-e^y\wedge e^{\beta}\bigg),\nn \\
F_3&=&-MKK^{\prime}F(y)\f{df(\tau)}{d\tau} e^{\psi}\wedge\bigg(e^{\theta}\wedge e^{\phi}-e^y\wedge e^{\beta}\bigg).
\eeqn
In order to be consistent with the Bianchi identity of these 3-form 
field strengths we shall set $\f{df(\tau)}{d\tau}=1$ and the quantization 
condition of $F_3$ imply \cite{ehk}
\be
K^{\prime}=4\pi^2\alpha^{\prime}, \quad K=\f{9}{8\pi^2} (p^2-q^2).
\ee

The self dual five-form field strength is
\beqn
& &{\tilde F}_5=-\f{h^{-2}}{g_s}\f{\p h}{\p \tau}d\tau\w dx^0\w\cdots\w dx^3-\f{h^{-2}}{g_s}\f{\p h}{\p y}dy\w dx^0\w\cdots\w dx^3\nn \\& &
- \f{1}{g_s}\f{\p h}{\p \tau} G^4_2 e^{\theta}\w e^{\phi}\w e^y\w e^{\psi}
+\f{1}{g_s}\f{\p h}{\p y}\sqrt{wv} G^2_1 G^2_2 d\tau\w e^{\theta}\w e^{\phi}\w e^{\beta}\w e^{\psi}.\nn \\& &
\eeqn

It is easy to note that  the warp factor is a function of both the radial 
coordinate $\tau$ and the angular coordinate $y$, as we have written in the
expression of five form field strength, because of the appearance of $F(y)$
in the three form field strengths i.e. if we look at the Bianchi identity
or the equation of motion associated to  the five form field strength then 
the RHS of this equation, $H_3\w F_3$ depends on F(y).  
The explicit form of the warp factor $h(\tau, y)$  is 
\be
h=4 B\int_{\tau} \f{\tau d\tau }{G^4_2(\tau)}-{\tilde B}\int_{\tau}\f{d\tau}{G^2_2}+\f{1}{G^2_1G^2_2}H(y)+constant,
\ee
where ${\tilde B}$ is a constant of integration and 
$B=\f{g^2_s K^2 {K^{\prime}}^2}{2 (1-y_1)^2 (1-y_2)^2} $ and H(y) is
\be
\label{H(y)}
H(y)=-\f{(g_sMKK^{\prime})^2}{2(b-1)}\bigg(\f{1}{1-y}+\f{(1+2y_1)(1+2y_2)log(y_3-y)}{2 (b-1)}\bigg)+constants.
\ee
It is interesting to note that all the roots of $b-3y^2+2y^3$ appear in H(y) 
and more importantly, this function do not diverges for $y_1\leq y\leq y_2$
\cite{ehk}. Given the 10-dimensional metric ansatz in 
eq.(\ref{metric_10}) this form of
H(y) is universal in the sense that one will get  a term in the warp 
factor which depends only on the $y$ coordinate and whose form is that in 
eq.(\ref{H(y)}) even if the 6-dimensional Ricci flat metric is of the 
$ds^2=G^2_1(\tau)[d\tau^2+{e^{\psi}}^2]+G^2_2(\tau)
[{e^{\theta}}^2+{e^{\phi}}^2]+G^2_3(\tau){e^y}^2+ G^2_4(\tau){ e^{\beta}}^2$
form.

The tilde five form field strength integrated over the 
``5-cycle=$S^2\times S^3$'' is 
\be
\oint_{S^5}{\tilde F}_5=\f{1}{g_s}\bigg[4B\tau-const.+2\bigg(\f{G_2\f{dG_2}{d\tau}}{G^2_1}+\f{G^2_2\f{dG_1}{d\tau}}{G^3_1}\bigg)\bigg]\oint_{S^5} e^{\theta}\w e^{\phi}\w e^y\w e^{\beta}\w e^{\psi}\\
\ee

Let us now evaluate the warp factor $h(\rho,y)$ and $\oint_{S^5}{\tilde F}_5$
for eq.(\ref{solution})  and the results of it in terms of our radial 
coordinate $\rho$ are
\be
h(\rho,y)=\f{2}{3}B\int_{\rho} dx \f{log[x^6+3 s^2 x^4+3 s^4 x^2]}{[x^5+3 x^3 a^2+3 s^4 x]}+\f{H(y)}{(\rho^2+s^2)^2}+const,
\ee
where we have set the constant ${\tilde B}$ to zero. In the 
$\rho \rightarrow 0$ limit the warp factor has a piece which depends on 
$\rho$ logarithmically and a piece independent of $\rho$ but depends on the 
angular coordinate $y$ through $H(y)$  and 
\beqn
& &\oint_{S^5}{\tilde F}_5=\f{1}{g_s}\bigg[\f{2B}{3} 
log[\rho^6+3s^2\rho^4+3s^4\rho^2]-const.\nn \\& &+2\bigg(1+\f{(\rho^2+s^2)(\rho^6+3s^2\rho^4+3\rho^2s^4+3s^6)}{\rho^2(\rho^6+4 s^2\rho^4+6s^4\rho^2+3s^6)}\bigg)\bigg] \oint_{S^5} e^{\theta}\w e^{\phi}\w e^y\w e^{\beta}\w e^{\psi}.\nn \\& &
\eeqn
{\bf Note Added}: After submitting the paper to arXiv.org, we are informed by 
C.N. Pope of their paper  \cite{pc}. For $\Lambda=0, \lambda=6, n=2, 
\kappa=c_1-s^6$ and
taking the 4-dim  geometry as the base of the
Einstein-Sasaki metric of $Y^{p,q}$ i.e. ${e^{\theta}}^2+{e^{\phi}}^2+{e^y}^2+{e^{\beta}}^2$ along with a choice for the connection
gives rise to our metric.   
\section{Acknowledgment}
We would like to thank O. Aharony, M. Berkooz, Andrei and D. Gepner for
many useful discussions and suggestions on the subject also to C. Herzog and
C. Pope for useful correspondences. The financial help 
from Feinberg 
graduate school is gratefully acknowledged.
 

\end{document}